\documentclass[12pt]{iopart}
\usepackage{graphics}
\begin{document}

\title{Tunnel and thermal c-axis transport in BSCCO in the normal and pseudogap state.}

\author{M Giura\dag 
\footnote[4]{E-mail:maurizio.giura@roma1.infn.it.}, R
Fastampa\dag, S Sarti\dag, N Pompeo\ddag, E Silva\ddag}

\address{\dag\ Dipartimento di Fisica and Unit\`a CNISM,
Universit\`a ``La Sapienza'', Piazzale Aldo Moro 2, I-00185 Roma,
Italy}

\address{\ddag\ Dipartimento di Fisica ''E. Amaldi'' and Unit\`a CNISM,
Universit\`a Roma Tre, Via della Vasca Navale 84, I-00146 Roma,
Italy}

\begin{abstract}
We consider the problem of $c$-axis transport in double-layered
cuprates, in particular with reference to
Bi$_{2}$Sr$_{2}$CaCu$_{2}$O$_{8+\delta}$ compounds.  We exploit the effect
of the two barriers on the thermal and tunnel transport.  The
resulting model is able to describe accurately the normal state
$c$-axis resistivity in Bi$_{2}$Sr$_{2}$CaCu$_{2}$O$_{8+\delta}$, from the
underdoped side up to the strongly overdoped.  We extend the model,
without introducing additional parameters, in order to allow for the
decrease of the barrier when an external voltage bias is applied.  The
extended model is found to describe properly the $c$-axis resistivity
for small voltage bias above the pseudogap temperature $T^{*}$, the
$c$-axis resistivity for large voltage bias even below $T_c$, and the
differential $dI/dV$ curves taken in mesa structures.
\end{abstract}



\maketitle

\section{Introduction}
\label{intro}
The $c$-axis conductivity in layered superconductors is at the same
time a fascinating puzzle and a source of interesting information on
the nature of high-$T_{c}$ cuprates.  Focussing on
Bi$_2$Sr$_2$CaCu$_2$O$_{8+\delta}$ (Bi:2212), on passing from
underdoping to overdoping, the $c$-axis resistivity $\rho_c$ changes
from monotonically decreasing from $T_{c}$ to high temperature to a
non-monotonic behaviour, displaying a minimum at a temperature usually
higher than the pseudogap temperature $T^{*}$
\cite{watanabe,esposito,giuraPRB03}.  This latter feature has led to a
description in terms of tunnelling phenomena along the $c$-axis
accompanied by some kind of charge localization on the $ab$ planes
\cite{anderson5,kumar}.  Such localization becomes even stronger in
the so-called pseudogap phase \cite{timusk}, as testified by an
increase of $\rho_c$ and a change of slope in the in-plane
resistivity ($\rho_{ab}$) \cite{watanabe}.\\
In addition to the study of the $c$-axis resistivity, many
investigations have been devoted to the study of the nonlinear electrical
response exhibited by the $I-V$ characteristic
\cite{tanabePRB96,schlengaPRB98}, $dI/dV$ curves
\cite{rennerPRL98,suzukiPRL99,suzukiPRL00,krasnovPRB02}, $\rho_{c}$ at
high applied voltages \cite{latyshevPRL99} or a combination of such
measurements \cite{krasnovPRL00}.  While in general terms it is fair
to say that by means of such measurements one can get information 
both on the normal state underlying the pseudogap state and on the
pseudogap state itself as well as on the superconducting state
quasiparticle response, the discussion on such data is still
very active \cite{stajicPRB03}.\\
Confining for the moment our attention to some feature of the
measurements taken in artificially patterned mesa structures, the so
called ``Intrinsic Tunnelling Spectroscopy'', we recall some of the
results that are particularly pertinent to the present work.  In
overdoped and optimally doped BSCCO mesas
\cite{latyshevPRL99,krasnovPRL00}, composed by several unit cells, the
differential conductivity exhibits two distinct behaviours for
voltages higher than a characteristic value, $V>V_{g}$, and for small
(vanishing) voltages, $V\rightarrow 0$.  The ($V\rightarrow 0$)
behaviour usually regarded as originating from quasiparticles, but the
strangest features came from the large voltage behaviour indicated as
``normal'' resistance $R_{N}$.  In fact, it was found that $R_{N}$ was
linear in $T$ down to $T\rightarrow 0$, in contrast with the usual
behaviour of $\rho_c$.  This feature does not seem to have been
satisfactorily described.\\
In two recent papers \cite{giuraPRB03,giuraPRB04} we have addressed
the issue of the out-of-plane, normal state $\rho_c$ resistivities
within a phenomenological model for bilayer superconductors, based on
the existence of two energy barriers with different height and width,
over which the electrical transport was determined by two
complementary processes, namely incoherent tunnelling \cite{kumar} and thermal
activation across the two barriers.  In spite of the simplicity of the
model, we have been able to fit a very wide set of $\rho_c$ data taken 
in BSCCO following the change of the
temperature dependence of the $c$-axis resistivity with increasing
doping $\delta$.  More precisely we showed that:\\
\textit{(i)} For $T>T^*$ the model (which explicitly refers to the
genuine normal state, with an electron density of states that is
uniform in energy) was able to describe the $\rho_c$ data with a
reduced number of fitting parameters, with a direct physical
interpretation.  The accuracy of the fit was as good in the underdoped
region as in the overdoped one.\\
\textit{(ii)} For $T_c<T<T^*$, using the normal $c$-axis resistivity
as determined by
our model, we described the loss of conductivity in the pseudogap
state by a reduction of the effective number of charge carriers.
We obtained the fraction $\eta(T)$ of the ``gapped'' carriers
that do not participate to the conduction along the c-axis, defined 
through
\begin{equation}
\sigma_c=\sigma_{c,n}[1-\eta(T)] \rightarrow \rho_c(T)=\frac{\rho_{c,n}(T)}{1-\eta(T)}
\label{eta_def}
\end{equation}
and we
found that the curves $\eta(T)$ at different dopings could be scaled
onto a universal curve \cite{giuraPRB04,giuraM2S}.\\
In this paper we extend the analysis of the data by means of the 
proposed model, by extending it to nonzero voltages, thus allowing 
a comparison of the model with the data taken on the mesa structures.
The main result of this paper will be a comprehensive and satisfactory
explanation of the mesa resistivity branches, both for $V\rightarrow
0$ ($R_{0}$) and for $V>V_{g}$ ($R_{N}$).\\
The paper is organized as follows.  In the next Section, we introduce
the basic idea of the two-barrier model. We extend the model to
finite applied voltage, and we discuss the limiting forms. In
Sec.\ref{fits} we compare the model to a variety of measurements taken
at various $T$ on BSCCO crystals.  Conclusions are reported in
Sec.\ref{conc}

\section{Two-barrier model}
\label{model}
\begin{figure}[htb] 
\begin{center}
     \includegraphics{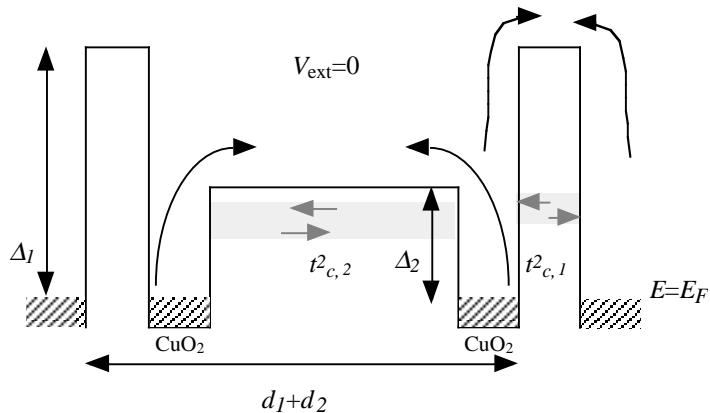}
\end{center}
\caption{Sketch of the energy profile used to develop the model for 
the small voltage $c$-axis resistivity. The different processes 
(thermal activation and tunnelling) and heights $\Delta_{i}$ and widths 
$d_{i}$ of the 
barriers are depicted. The barriers can be thought of as arising from the 
Ca layer and the Sr-Bi-Bi-Sr block, respectively.}
\label{rectbarrier}
\end{figure}
As already mentioned, in two recent papers \cite{giuraPRB03,giuraPRB04} we have presented a
phenomenological model for the out-of-plane resistivity $\rho_{c}$ in
bilayer superconductors.  In this model the temperature dependence of
the $c$-axis resistivity in the genuine normal state ($T>T^{*}$, where
$T^{*}$ indicates the pseudogap opening) arises from the existence of
two different energy barriers along the $c$-axis, taken of 
rectangular shape with different
heights ($\Delta_{1}$ and $\Delta_{2}$, as measured from the Fermi 
level) and widths $d_{1}$ and $d_{2}$, as depicted in
Fig.\ref{rectbarrier}.  In this spatial energy landscape, in the
normal state two mechanisms (for each barrier) contribute to the
interlayer transport: \textit{(i)} thermal activation of the carriers
above the barrier \textit{(ii)} tunnelling of the charge carriers
partially inhibited by incoherent in-plane scattering.\\
The tunnelling contribution over each of the barriers was identified as
given by the model of Ref.\cite{kumar}, for which the tunnelling
contribution to the $c$-axis resistivity is modified by the existence
of the interplanar incoherent scattering, leading to a confinement of
the charge carriers in the $a,b$ planes.  In this model, the in-plane
phononic, inelastic scattering, with characteristic time $\tau$, reduces the
tunnelling matrix element $t_{c}$ to $(2\tau/\hbar)t_{c}^{2}<<t_{c}$
when $\tau$ is much smaller than the time between two tunnelling
elements.  These features of the model imply a relation between the in-plane and the
out-of-plane resistivities.  The overall expression for the normal
state effective resistivity along the c-axis,
$\rho_{c,n}$, is obtained as a function of the in-plane $\rho_{ab,n}$
from the series of the two barriers \cite{giuraPRB03}:
\begin{eqnarray}
\rho_{c,n}& = & \frac{1}{d_{1}+d_{2}}\sum_{i=1,2} \left[ \frac{t_{c0,i}^2}
{a\rho_{ab,n}+b}+\beta e^{-\Delta_i/k_BT} \right]^{-1} \nonumber\\
 & = & \sum_{i=1,2}\frac{d_{i}}{d_{1}+d_{2}}\left[
 \frac{1}{\rho_{tun,i}}+ \frac{1}{\rho_{th,i}}\right]^{-1}
\label{rholin}
\end{eqnarray}
where the last equality defines the tunnelling ($\rho_{tun,i}$) and
thermal ($\rho_{th,i}$) resistivities of the $i-$th barrier,
respectively, the term $a\rho_{ab,n}+b$ takes into account the
increased localization due to the in-plane phononic scattering and
$d_i$ ($i=1,2$) is the spacing between the layers (we have taken
$d_1=3$ \AA, $d_2=12$ \AA, the bilayer and the Sr-Bi-Bi-Sr block
thickness, respectively).  In this expression
$t_{c0,i}^2\propto\exp[-2d_i\sqrt{2m^{*}\Delta_i}/\hbar]$ (with
$m^{*}=4.6 m_{e}$ \cite{poole}) and $\Delta_i$ is the height of the
$i$-th barrier.  Since the in-plane resistivity $\rho_{ab,n}$ is
independently measured, Eq.\ref{rholin} contains five adjustable
parameters: $a, b, \beta, \Delta_{1}, \Delta_{2}$, with the additional
constraint that $\Delta_{1}-\Delta_{2}$ is doping-independent.
$t_{c0,i}^2$ is determined by $d_{i}, \Delta_{i}, m^{*}$, with its
prefactor absorbed in $a$ and $b$.  Note that $a, b, \beta$ can be
combined in such a way that only two of them determine the shape of
the fitting curve, while the third affects the fit only through a
scale factor.\\
\begin{figure} [htb]
\begin{center}
      \includegraphics{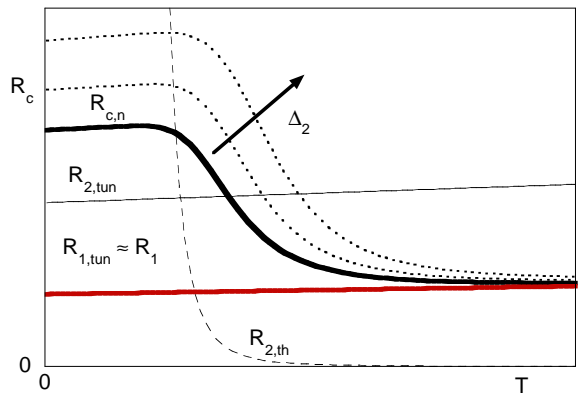}
\end{center}
\caption{General overview of the temperature dependence of the
$c$-axis resistance as calculated in the two-barrier model (black
thick line), $R_{c,n}=\rho_{c,n}\frac{d_{1}+d_{2}}{S}$, where S is the
transverse section.  Here we have neglected (see text) the thermal
conductivity across the higher barrier.  The contributions of the
first barrier (red thick line) and the thermal (dashed line) and
tunnel (thin continuous line) contributions of the second barrier are
separately plotted ($R_{i,tun}=\rho_{i,tun}\frac{d_{i}}{S}$ for the
tunnelling resistance of each barrier, and similarly for the thermal
contribution).  We also report the effect of increasing $\Delta_{2}$
(dotted lines).}
\label{rhomaster}
\end{figure}
In order to avoid unnecessary complications, we recall that the thermal
contribution to the higher barrier (1) was found to be negligible
\cite{giuraPRB03}, with $\Delta_{1}/k_{B}>2000 K$, so that for this
barrier we can safely assume $\rho_{1}\sim\rho_{tun,1}$ in the whole
temperature range explored ($T<$300K). The reference expression for 
the effective resistivity then simplifies to:
\begin{equation}
\rho_{c,n}=\frac{1}{d_{1}+d_{2}}\left(\rho_{1,tun}d_{1}+
\frac{\rho_{2,tun}\rho_{2,th}}{\rho_{2,tun}+\rho_{2,th}}d_{2}\right)
\label{overall_rho_c}
\end{equation}
In Figure \ref{rhomaster} we present a sketch of the resulting
temperature dependence of the $c$-axis normal state resistivity
$\rho_{c,n}$.  A thorough discussion of the parameters involved in
equation \ref{rholin}, and of their dependence on doping $\delta$ has
been given in a previous paper \cite{giuraPRB03}. However, it is 
interesting to plot separately the various contributions to the 
overall resistivity, $\rho_{tun,i}$ and $\rho_{th,i}$.\\
In particular, equation \ref{overall_rho_c} predicts for $\rho_{c,n}$
the following general behaviour: at very low $T$, $\rho_{2,th}\gg
\rho_{2,tun}$, so that $\rho_{c,n}$ is substantially the sum of the
two tunnelling resistivities $\rho_{1,tun}$ and $\rho_{2,tun}$, both
of them linearly increasing as a function of $T$.  As a result,
$\rho_{c,n}$ increases linearly with $T$ at low temperatures.  At
intermediate $T$, $\rho_{2,th}< \rho_{2,tun}$ so that in this
temperature region the resistivity of the barrier 2 is dominated by
the thermal activation term and is thus decreasing with $T$, and so is
also the overall $\rho_{c,n}$.  Finally, at high $T$
$\rho_{2,th}\rightarrow 0$, so that $\rho_{c,n}\simeq \rho_{1,tun}$,
again increasing linearly with $T$.  The crossover from thermally
activated ($\rho_{c,n}\sim\rho_{2,th}$) to tunnel dominated behaviour
($\rho_{c,n}\sim\rho_{1,tun}$) occurs for $\Delta_{2}/k_{B}T\sim$1.
It is interesting to note that this crossover reproduces in a natural
way the minimum experimentally found in measurements of $\rho_{c}$ in
optimal and overdoped samples, assigning it to a smaller $\Delta_{2}$
with respect to underdoped samples.  In this framework, the minimum is
not a manifestation of the opening of the pseudogap, which occurs
at a lower temperature.  We stress that, as mentioned in
\cite{giuraPRB03}, the fits of $\rho_{c}$ (at low voltages) only allow
us to set a lower bound for $\Delta_{1}-\Delta_{2}$.\\
A second interesting feature of the model is that it naturally leads
to $c$-axis resistance linear in $T$ at low temperature, as observed
in mesa structures at high voltage bias (that is for applied voltages
$V>V_{g}$, where $V_{g}$ is a typical voltage applied to the mesa
structure above which the $I-V$ curve becomes approximately linear).
These data are reported, e.g., in Fig.2 of Ref.\cite{latyshevPRL99}
and in Fig.1 (inset) of Ref.\cite{krasnovPRL00}.\\
Within our model, as the voltage is increased the effective height of
the second barrier decreases (see also below) while the much higher
first barrier remains substantially unaffected.  As a result, both the
thermal and tunnelling resistivities of the second barrier become much
smaller than their zero voltage values, so that the high voltage
resistivity approaches the tunnelling value $\rho_{1,tun}$:
\begin{equation}
R_N=R(V>V_g)\simeq R_{1,tun}= 
\frac{a\rho_{ab}+b}{t_{c1}^{2}}\frac{1}{S}
\label{RlinT}
\end{equation}
($S$ is the cross-section of the mesa structure) which, due to the
linear dependence (as a function of $T$) of the in-plane resistivity
directly explains the (so far unexplained) linear behaviour of $R_N$,
as observed in mesa structure data \cite{latyshevPRL99,krasnovPRL00}.\\
In the following, we analytically extend our model to nonvanishing
voltages $V$, in order to compare the predictions with the large
amount of data taken on mesa structures.\\
To be quantitatively applied to measurements of the nonlinear
resistance, $I-V$ and $dI/dV$ characteristics, the model requires a
generalization that includes a direct effect of the applied voltage on
the barrier height.  In this case the rectangular barrier framework is
no longer appropriate and a more accurate modelling of the barrier
must be performed in order to take into account the decreasing of the
barrier height with the external electric field.  As depicted in
Fig.\ref{parabolic}, we have chosen parabolic barriers since this
choice allows for analytical results \footnote{Many other choices are
clearly possible.  We mention that we have performed the calculations
and fittings also with trapezoidal barriers, obtaining the same
results as for the parabolic case, with slightly varied
numerical factors in the parameters.}.  The energy profile along the
$c$-axis ($x$ direction) is then given by two parabolic barriers with
the shape: $U_{i}(x)=U_{ci}-\frac{\alpha}{2}(x-\frac{d_{i}}{2})^2$ in
the interval $0<x<d_{i}$, and zero otherwise.  The parameter $U_{c,i}$
is directly related to the previously defined height of the barrier
$\Delta_i$ through the relation $U_{c,i} = \Delta_i+E_F$ (see also
figure \ref{parabolic}), while $\alpha=8U_{ci}/d_{i}^{2}$.  We stress
that the choice of parabolic barriers, given the same widths $d_{i}$
as in the rectangular barriers, does not introduce additional
parameters.\\
\begin{figure}[htb] 
\begin{center}
     \includegraphics{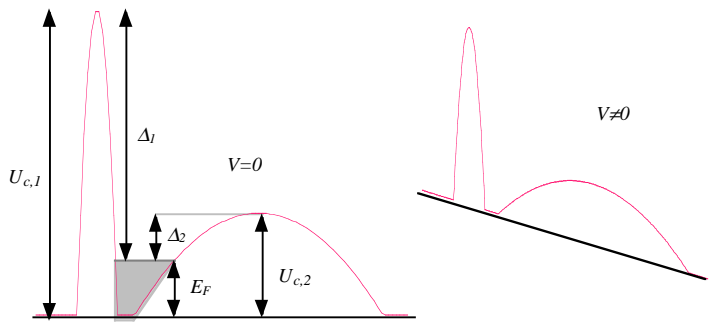}
\end{center}
\caption{Left panel: generalized two-barrier model for the study of linear and 
nonlinear resistivity. The parabolic shape of the barrier is one of 
the possible choices that have to be selected, in 
order to incorporate in the model the reduction of the barrier height 
due to an external bias (finite voltage). The effect of the external 
voltage is sketched in the right panel.}
\label{parabolic}
\end{figure}
Now we describe the different transport processes in the finite
voltage regime.  We consider first the activation process.  In the
presence of an external electric field $\mathcal{E}$, the barrier
becomes $U_{i}^{\prime}(x)=U_{i}(x)-e\mathcal{E}x$.  Assuming that the
volume density of carrier $n=const$ outside the barrier, the thermal
charge current density is \cite{ma}:
\begin{equation}
    J_{th,i}=-\frac{eDn\left(1-e^{eV_{i}/k_{B}T}\right)}
    {\int_{x_{i,1}}^{x_{i,2}}e^{\left[U_{i}^{\prime}(x)-E_{F}\right]/k_{B}T}dx}
\label{Jth}
\end{equation}
where $D$ is the diffusion coefficient and $V_{i}$ is the voltage drop 
across the $i$-th barrier. Thus, one has:
\begin{equation}
    \int_{x_{i,1}}^{x_{i,2}}e^{\left[U_{i}^{\prime}(x)-E_{F}\right]/k_{B}T}dx=
    e^{\Delta_{i}/k_{B}T} \int_{x_{i,1}}^{x_{i,2}}
    e^{-\left[\frac{\alpha}{2}(x-\frac{d_{i}}{2})^2+e\mathcal{E}x\right]/k_{B}T}dx
\label{integral}
\end{equation}
where $\mathcal{E}$ is the electric field.  The integral on RHS can be
expressed in terms of the error function
$\Phi(x)=\frac{2}{\pi}\int_{0}^{x}e^{-t^{2}}dt$, and after some
straightforward calculation one gets the contribution of the
activation to the diffusion transport current for a single barrier:
\begin{equation}
    J_{th,i}=-\sqrt{\frac{16U_{ci}}{\pi k_{B}T}}\frac 
    {eDn\left(1-e^{eV_{i}/k_{B}T}\right) \sinh (eV_{i}/2k_{B}T)} 
    {d_{i}
    \exp\left[\frac{(eV_{i})^{2}}{16U_{ci}k_{B}T}\right]
    2 \Phi\left(\sqrt{\frac{(4U_{ci}-eV_{i})^{2}-16E_{F}U_{ci}}{16U_{ci}k_{B}T}}\right)}
\label{Jthfin}
\end{equation}
In considering the tunnel resistivity, we make use of the
expression
\begin{equation}
    J_{i}(V_{i})=-
    \int_{0}^{\Delta_{i}+E_{F}} 
    \frac {t_{ci}^{2}(E)}{e(a\rho_{ab}+b)}
    \left[
    f(E)-f(E+eV_{i})
    \right] dE
\label{Jtun}
\end{equation}
as for the field emission effect with the additional localization
considered in reference \cite{kumar}, which we assume to be valid also
in the nonlinear regime.  Here $t_{ci}^2(E)$ is the (energy dependent)
transparency of the barrier and $f$ is the Fermi function.  To
calculate $t_{ci}^2(E)$, we use the same barrier's profile $U(x)$ used
to calculate the thermal contribution.  For a generic energy level $E$
in an external field the barrier's profile is shown in
Fig.\ref{parabolic}.  After some calculation
\cite{abramowitz,gradshteyn}, one has:
\begin{eqnarray}
t_{ci}^{2}(E) & = & \exp \left\{-\frac{2\sqrt{2m^{*}}}{\hbar} 
	\int_{x_{i,1}}^{x_{i,2}} \left[\frac{\alpha}{2}x^{2}+
	\left(\frac{\alpha d_{i}}{2}-e\mathcal{E}_{i}\right) x-E\right]^{1/2} dx
	\right\} \nonumber\\
	 & = &  \exp\left\{
     -\frac{\pi\sqrt{2m^{*}U_{c,i}}d_{i}}{4\hbar}
     \left[
     \left(1-\frac{eV_{i}}{4U_{c,i}}\right)^{2}-\frac{E}{U_{c,i}}
     \right]\right\}
\label{tcnonlinear}
\end{eqnarray}
for $E<U_{ci}\left(1-\frac{eV_{i}}{4U_{ci}}\right)^{2}$ and
$t_{ci}^2=1$ otherwise.  Here $\mathcal{E}_{i}$ is the electric field
on the $i$-th barrier.
The total current in the series of the two barriers is the sum of the
various contributions each with its specific values of the
parameters.\\
It must be noted that, in the nonlinear regime, several complications
arise.  First of all, the energy dependences of the current densities
(thermal and tunnelling) is very complex by itself, and additionally
due to the very similarity of the energy scales involved: the energy
$E$, the Fermi level $E_{F}$ and, in some ranges, the thermal energy
$k_{B}T$ are all of the same order of magnutude.  A second, more
subtle complication is due to the nature of the double barrier: the
experimentally controlled parameter is the external potential $V$,
applied at the series of two barriers (assuming, as in measurements of
the mesa structures, that one knows the number of junctions $N$
involved, the applied voltage is just $NV$).  However, $V$ is
\textit{not} the voltage across each barrier, but the sum of the
potential drops over the two barriers: $V=V_{1}+V_{2}$.  Due to
the temperature and to nonlinear effects, the ratio $V_{1}/V_{2}$ is
not constant.  One can implicitly obtain each $V_{i}$ by requiring
that the total current through the two barriers be the same:
$J=J_{1}=J_{2}$, with $J_{i}=J_{th,i}+J_{tun,i}$ (the cross section
$S$ is the same throughout the sample).  One can thus solve
numerically to obtain $V_{1}$ and $V_{2}$ as a function of $V$ and
the $I(T,V)=JS$ and $R_{c}(T,V)=V/I$ for every temperature
and arbitrary $V$.\\
We are now in the position to perform some comparison between the 
model here developed and the data for the $c$-axis transport. The next 
Section is dedicated to this task.

\section{Comparison with experiments}
\label{fits}
In this Section we report the application of the model developed in
Sec.\ref{model}, based on parabolic barriers, to experimental data for
$c$-axis transport in BSCCO reported in the literature or measured in
our laboratory.  We first analyze the data for the $c$-axis resistance
taken in optimally doped BSCCO single crystal mesa structures
\cite{krasnovPRL00}.  Since those data were taken at small
($V\rightarrow 0$) and large ($V=V_{g}$) voltages, the simultaneous
fitting of both sets of data allows a precise determination of the
doping-independent value of $\Delta_{1}-\Delta_{2}$.  Once determined,
this value will be kept fixed in the fits of all the other data at
various dopings.\\
\begin{figure} [htb]
\begin{center}
      \includegraphics{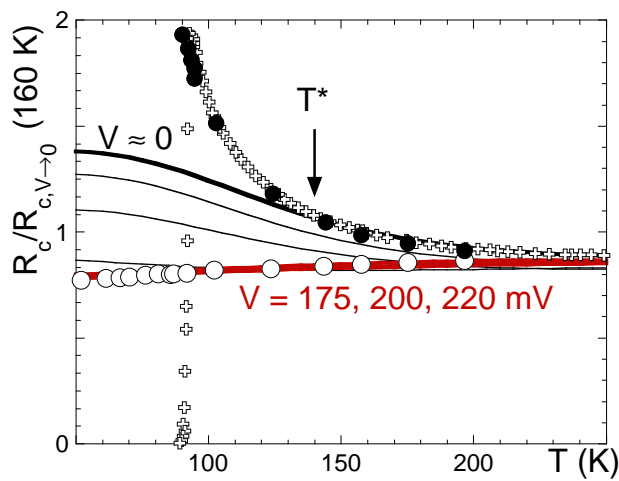}
\end{center}
\caption{Normalized data for the $c$ axis resistivity as reported in
\cite{krasnovPRL00} at high voltage, $R_{c,V_{g}}$, open circles, and
small voltage, $R_{c,V\rightarrow 0}$, full dots, compared to our data
for in a BSCCO single crystal with nominal $\delta=$0.255 (crosses).
Continuous curves show the voltage bias dependence of $R_{c}$,
calculated on the basis of the model described in Sec.\ref{model} with
$\Delta_{2}/k_{B}=$690 K: thick black line, $V\simeq 0$; thick red
lines: $V=$175, 200, 220 mV; thin lines, from top to bottom: 45, 90,
160 mV.}
\label{figV0Vg}
\end{figure}
Due to the illustrative nature of these fittings, we describe in
some detail the procedure that we followed in order to reduce the
number of fitting parameters.  For the small voltage resistivity, we
have to refer to Eq.\ref{rholin}, which contains the in-plane
resistivity $\rho_{ab}$.  Since those data were not available in
\cite{krasnovPRL00}, we compared the out-of-plane resistance with our
data taken in BSCCO single crystals at various doping $\delta$ by an
eight terminal contact method, which allows us to extract
simultaneously $\rho_{c}$ and $\rho_{ab}$ \cite{esposito}.  As
reported in Fig.\ref{figV0Vg}, our data for a crystal with nominal
doping $\delta=$0.255 coincides with the $R_{c,V\rightarrow 0}$ data
in \cite{krasnovPRL00} within an overall scale factor.  We then
decided to use our data for $\rho_{ab}$ in the fitting.  By adjusting
$\Delta_{2}$, $\beta$, $a$ and $b$ it is possible to fit well the
normal state $R_{c,V\rightarrow 0}$ down to a temperature $T^{*}$
where the pseudogap opens (we come back to this point below, see also
\cite{giuraPRB03,giuraPRB04,giuraM2S}). For the data of $R_{c,V_{g}}$
one has to refer to the complex procedure decribed in Sec.\ref{model},
calculating the total $J$ with the dynamic voltage partition between
the two barriers.  In this case the fit is sensitive to $\Delta_{1}$.
In Eq.s \ref{Jth},\ref{integral},\ref{Jthfin},\ref{Jtun} we have used
$E_{F}/k_{B}=$1000 K \cite{poole}.  By simultaneously fitting
$R_{c,V\rightarrow 0}$ and $R_{c,V_{g}}$ a strong constraint is put on
the parameters.  We found that it was indeed possible to fit
\textit{both} $R_{c,V\rightarrow 0}$ and $R_{c,V_{g}}$ (a nontrivial
result) with the choice $(\Delta_{1}-\Delta_{2})/k_{B}=$2000 K, of the
same order of magnitude but smaller than the value $\sim$5000 K
estimated in our previous works.  We remark that no additional or
hidden parameters are required, besides those above mentioned.  In
Fig.  \ref{figV0Vg} we report the evolution of the nonlinear $R_{c}$
at various voltages.  It is seen that the fitting of $R_{c,V_{g}}$ is
obtained with $V$=180 mV, as compared to $V_{g}\simeq$150 mV reported
in \cite{krasnovPRL00}.\\
As a second illustration of the applicability of the model, we present
in Fig.  \ref{figrc} the fits of the small voltage resistivity,
$\rho_{c,V\rightarrow 0}$, as measured by us in BSCCO single crystals.
In these fits we have taken $(\Delta_{1}-\Delta_{2})/k_{B}=$2000 K, as indicated
by the simultaneous fittings above performed.
Thus, once a simultaneous fit of low and high voltage resistance has 
been performed at a single doping, the fits at all the other dopings 
contain only four adjustable parameters (one of which acts as a scale 
factor only, see Sec.\ref{model}).
As can be seen, the
fits capture the minimum of $\rho_{c}$.  The behaviour
of the parameters as a function of the doping has been already
discussed \cite{giuraPRB03}.  Here, we recall that the fits depart
from the data when the pseudogap opens.  In this case from Eq.
\ref{eta_def} one can evaluate the fraction $\eta$ of the carriers
that no more participate to the conduction below $T^{*}$.  It is not
the purpose of this paper to discuss the behaviour of $\eta$, that has
been partially discussed before \cite{giuraPRB04} and elsewhere
\cite{giuraM2S}.  However, for completeness we report some of the
resulting $\eta$ in the inset of Fig. \ref{figrc}.\\
\begin{figure} [htb]
\begin{center}
      \includegraphics{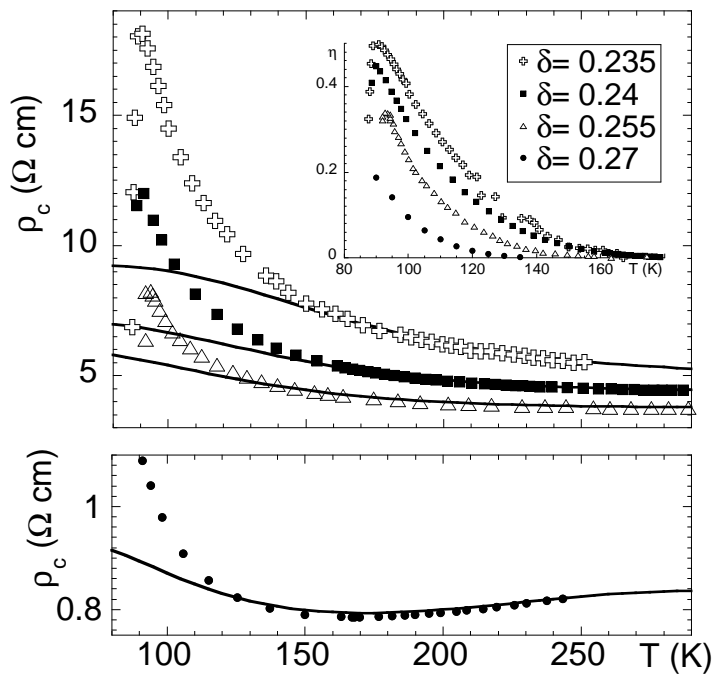}
\end{center}
\caption{Typical fits of the $c$ axis resistivity at small voltage
with the proposed model, continuous lines, compared to our data in
BSCCO single crystals at various dopings (symbols).  The data for the
most overdoped sample are reported in the lower panel with a much
enlarged scale to show the minimum in $\rho_{c}$.  In the inset we
report the fraction of charge carriers that do not participate to the
conduction after the pseudogap opening.}
\label{figrc}
\end{figure}
As a third and final illustration we calculate the differential
$dI/dV$ curves by means of the nonlinear equations
\ref{Jth},\ref{integral},\ref{Jthfin},\ref{Jtun} and we compare the
result with the data measured in BSCCO mesa structures in
\cite{suzukiPRL00}.  The discussion of $dI/dV$ data is of particular
interest, since the ``dip and hump'' structure has been the subject of
many discussions
\cite{rennerPRL98,suzukiPRL00,tallonPRL99,rennerPRL99,krasnovPRL01}.
In particular, the survival of this structure at high temperatures is
not yet unanimously attributed to some specific mechanism, and it is
often interpreted as a persistence of the pseudogap at high
temperatures \cite{krasnovPRL05}.\\
Calculations of the $dI/dV vs. V$ curves using our model are reported 
in Fig. \ref{figdIdV}. In order to 
compare the calculations with the data in \cite{suzukiPRL00}, the curves have 
been shifted vertically. It is clearly seen that, within a single scale 
factor of order 1, our calculation captures reasonably well the 
experimental behaviour, at all temperatures. We stress that, once a 
single parameter set has been chosen for one of the curves, the others 
are calculated without further adjustments. The agreement is 
fairly good, especially taking into account the very simple (albeit 
analytically cumbersome) model.\\
\begin{figure} [htb]
\begin{center}
      \includegraphics{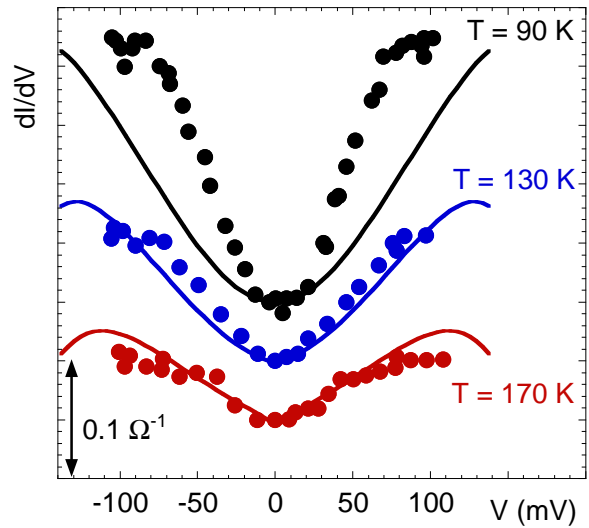}
\end{center}
\caption{Differential conductance $dI/dV vs.  V$ in a 
sample with $\delta=$0.25;  symbols: experimental data after
\cite{suzukiPRL00}; continuos lines: fits by means of the model described in
Sec.\ref{model}.  The calculated curves have been scaled by a single
overall factor.  The curves have been shifted vertically for clarity.}
\label{figdIdV}
\end{figure}
\section{Conclusions}
\label{conc}
In this paper we have reviewed the two-barrier model for the electric
transport along the $c$-axis in the double-layered superconductor
BSCCO. We have extended the model to nonvanishing voltage bias, in
order to compare the calculation with the measurements of the $c$-axis
resistance taken in mesa structures.  These measurements show rich and
peculiar behaviour, namely the minimum in $\rho_{c}(T)$ at some
doping, the crossover from concave upward $R_{c,V\rightarrow 0}(T)$ to
linear $R_{c,V_{g}}$ at high voltage bias, the shape of the $dI/dV vs.
V$ curves above $T_{c}$.  Nevertheless, quantitative fits to all these
behaviours may be obtained using a simple extension of our earlier
model.  It seems that double-layered tunnelling might be responsible
or co-responsible for many of the experimental observations.
\ack{We are grateful to E. L. Wolf and G. Gu for supplying the BSCCO 
crystals whose data are reported in Fig.\ref{figrc}.}

\end{document}